\documentclass[12pt,a4paper]{article}
\usepackage{latexsym,amsfonts}
\usepackage{exscale}
\usepackage{a4}
\usepackage[dvips]{graphicx} 
\usepackage{cite}

\newcommand{\beq}{\begin{equation}}
\newcommand{\eeq}{\end{equation}}
\newcommand{\beqa}{\begin{eqnarray}}
\newcommand{\eeqa}{\end{eqnarray}}
\newcommand{\te}{\theta}
\newcommand{\teb}{\bar{\theta}}
\newcommand{\teh}{\hat\theta}

\newcommand{\tehb}{\,\bar{\!\hat\theta}}

\newcommand{\cab}{C^{\alpha\beta}}
\newcommand{\cij}{C^{ij}}

\begin{document}

\thispagestyle{empty}


\vspace{4em}
\begin{center}

{\Large{\textbf{Seiberg-Witten Map for Superfields on $N\!=\!(\frac{1}{2},0)$ and 
$N\!=\!(\frac{1}{2},\frac{1}{2})$ Deformed Superspace}}}

\vskip 4em

{{\textbf{D\v zo Mikulovi\' c\footnote{dzo@theorie.physik.uni-muenchen.de}} }}

\vskip 2em

Universit\"at M\"unchen, Fakult\"at f\"ur Physik\\
Theresienstr.\ 37, D-80333 M\"unchen, Germany\\[1em]

\end{center}

\vspace{4em}

\begin{abstract}
\noindent
In this paper we construct the Seiberg-Witten maps for superfields on the 
$\te-\te$ deformed superspaces with $N\!=\!(\frac{1}{2},0)$ and 
$N\!=\!(\frac{1}{2},\frac{1}{2})$ supersymmety. We show that on the 
$N\!=\!(\frac{1}{2},0)$ deformed superspace there is no Seiberg-Witten map for 
antichiral superfields which is at the same time antichiral, local and which 
preserves the $N\!=\!(\frac{1}{2},0)$ supersymmetry. Solutions which break 
this requirements are presented.
On the $N\!=\!(\frac{1}{2},\frac{1}{2})$ deformed superspace we show 
that for the chiral gauge parameter, and therefore also for the chiral matter
field, there is no chiral Seiberg-Witten map. Some other possible 
Seiberg-Witten maps for the superfields are presented.
\end{abstract}
\newpage
\setcounter{page}{1}

\tableofcontents

\section{Introduction}

Field theories on canonically deformed space 
\beq
[\hat x^{i} ,\hat x^{j} ] = i \te^{ij} \, ,\qquad \te^{ij}=-\vartheta^{ji} \in 
\mathbb{R} \, ,
\eeq
have recently attracted much attention (for reviews and an exhaustive list of 
references see \cite{Konechny:0012,Douglas:0106,Szabo:0109}), mainly due to the
discovery of this noncommutative space in string theory 
\cite{Connes:1998,Douglas,Seiberg}. Based on the existence of different 
regularization procedures in string theory, Seiberg and Witten claimed in 
\cite{Seiberg} that certain noncommutative gauge theories are equivalent to 
commutative ones. In particular, they argued that there exists a map from a 
commutative gauge field to a noncommutative one, which is compatible with the 
gauge structure of each. This map has become known as the Seiberg-Witten map.

In \cite{Madore:0012,Jurco:0006,Jurco:0102,Jurco:0104} gauge theory on 
noncommutative space was formulated using the Seiberg-Witten map. In contrast 
to earlier approaches \cite{Armoni:0005,Bonora:0006,Bars:0103,Chaichian:0107},
this method works for arbitrary gauge groups. Using this method the problems 
of charge quantization \cite{Hayakawa:99121,Hayakawa:99122} and tensor product 
of gauge groups \cite{Chaichian:0107} were solved and  the standard model and 
GUT's were formulated at the tree level on noncommutative space 
\cite{Calmet:0111,Aschieri:0205}.   

Non(anti)commutative superspaces naturally arise in string theory as well with
$x-x$ deformation (canonical deformation) \cite{Chu:1999}, $\te-\te$ 
deformation \cite{Ooguri:0302,Ooguri:0303,Seiberg:0305} and $x-\te$ deformation
\cite{deBoer:0302}. General deformed superspaces were first studied  more 
closely in \cite{Ferrara,Klemm} and recently in \cite{Ferrara:0307} and in 
connection with the supermatrix model in \cite{Hatsuda:0306,Park:0307}. 

Since the work of Seiberg \cite{Seiberg:0305}, various aspects of field theory 
were considered on the $\te-\te$ deformed superspace with 
$N\!=\!(\frac{1}{2},0)$ supersymmetry. Renormalization properties of the 
Wess-Zumino model and gauge theories were considered in 
\cite{Britto:0306,Britto:0307,Grisaru:0307,Britto:0307:2,Romagnoni:0307,
Lunin:0307,Levell:0308,Berenstein:0308}.
Solitons, instanton solutions and some nonpertubative aspects were considered 
in \cite{Abbaspur:0308,Imaanpur:0308,Imaanpur:0311,Grassi:0311,Britto:0311,
Billo:0402}.
The generalization to $N=2$ and other interesting features have been explored 
in \cite{Berkovits:0306,Terashima:0306,Araki:0307,Chaichian:0307,Ivanov:0308,
Ferrara:0308,Alishahiha:0309,Sako:0309,Chandrasekhar:0310,Araki:0401,
Saemann:0401,Inami:0402,Imaanpur:0403}.

Till now, on the $\te-\te$ deformed superspace gauge theories were defined 
without Seiberg-Witten map. Similar to the nonsupersymmetric case, without the 
Seiberg-Witten map only gauge theories for the $U(N)$ gauge group can be 
formulated \cite{Terashima:0306,Britto:0311}. Hence, in order to consider 
arbitrary gauge groups we are forced to determine the Seiberg-Witten maps for 
superfields. 

In \cite{Mikulovic:0310} it was shown, that on canonically deformed $N\!=\!1$, 
$d\!=\!4$ Euclidean superspace there exist a local, chiral and supersymmetric 
Seiberg-Witten map for chiral superfields if we take the noncommutativity 
parameter to be selfdual. If the noncommutativity parameter is antiselfdual,
there exist a local, antichiral and supersymmetric Seibeerg-Witten map for 
antichiral superfields. Furthermore, it was shown that on the canonically 
deformed $N=1$, $d=4$ Minkowski superspace the Seiberg-Witten map is not 
compatible with locality, (anti)chirality and supersymmetry at the same time.  

It is an interesting question, if we face the same problems on the $\te-\te$ 
deformed superspace. The aim of this paper is to answer this question. First
we will recapitulate some well known properties of the $N\!=\!(\frac{1}{2},0)$ 
deformed Euclidean superspace \cite{Seiberg:0305} and construct the 
Seiberg-Witten maps in terms of component fields just to get some feeling for
the problems which will occur. Thereafter we will construct the Seiberg-Witten 
maps in terms of superfields and discuss briefly their properties. Finally we 
make the same considerations on the $N\!=\!(\frac{1}{2},\frac{1}{2})$
deformed superspace \cite{Ferrara:0307,Saemann:0401}.
We use the conventions of \cite{WessBagger}.

\section{$N\!=\!(\frac{1}{2},0)$ deformed Euclidean superspace}

We consider the following deformed superspace \cite{Seiberg:0305}:
\beqa \label{eq201}
&& [\hat x^{i} ,\hat x^{j} ] = \teb\teb\cij \, , \nonumber \\
\hat\mathcal{R}:
&&  [\hat x^{i} , \teh^{\alpha} ] = i \cab \sigma^{m}_{\beta\dot\beta} 
\!\bar{\,\hat\theta^{\dot\beta}} \, , \quad [\hat x^{i} , 
\!\bar{\,\hat\theta^{\dot\beta}} ] = 0 \, , \\
&& \{ \teh^{\alpha} , \teh^{\beta} \} = \cab \, , \quad
\{ \!\bar{\,\hat\theta^{\dot\alpha}} , \!\bar{\,\hat\theta^{\dot\beta}} \} = 
\{ \teh^{\alpha} , \!\bar{\,\hat\theta^{\dot\alpha}} \} = 0 \nonumber \, ,
\eeqa
where $\cab$ is a constant, symmetric deformation parameter and
\beq \label{eq202}
\cij = \cab \varepsilon_{\beta\gamma}(\sigma^{ij})^{\,\,\,\gamma}_{\alpha}
\eeq
is selfdual. Useful identities are 
\beqa
\cab &=& \frac{1}{2}\, \varepsilon^{\alpha\gamma} 
(\sigma^{ij})^{\,\,\,\beta}_{\gamma} C_{ij}  \, , \label{eq203} \\
|C|^{2} &=& \cij C_{ij} ~=~ 4 \det C \, .\label{eq204}
\eeqa
Non(anti)commutativity is indicated by a hat. On this deformed 
superspace $\tehb$ is not the complex conjugate of $\teh$, which is possible 
only in Euclidean space. We will be working on Euclidean $\hat\mathbb{R}^{4}$,
but we will continue to use the Lorentzian signature notation.  

Using $y^{m}=x^{m} + i\te\sigma^{m}\teb$ we can accompany (\ref{eq201}) with
\beqa \label{eq205}
&& [\hat y^{i} ,\hat y^{j} ] = 0 \, , \nonumber \\
\hat\mathcal{R}:
&&  [\hat y^{i} , \teh^{\alpha} ] = 0 \, , \quad [\hat x^{i} , 
\!\bar{\,\hat\theta^{\dot\beta}} ] = 0 \, , \\
&& \{ \teh^{\alpha} , \teh^{\beta} \} = \cab \, , \quad
\{ \!\bar{\,\hat\theta^{\dot\alpha}} , \!\bar{\,\hat\theta^{\dot\beta}} \} = 
\{ \teh^{\alpha} , \!\bar{\,\hat\theta^{\dot\alpha}} \} = 0 \nonumber \, .
\eeqa
Thus it is obvious to consider the deformed chiral superspace (\ref{eq205})
instead of (\ref{eq201}). Noncommutative functions and fields are defined as 
elements of the noncommutative algebra 
\beq \label{eq206}
\hat \mathcal{A} = \frac{\mathbb{C} \big[ [\hat y^{i},\teh_{\alpha},
\tehb_{\dot\alpha}] \big]}{I_{\hat\mathcal{R}}} \, ,      
\eeq
where $I_{\hat\mathcal{R}}$ is the two-sided ideal created by the relations 
(\ref{eq205}). 

The derivatives act on the coordinates as in the classical case 
\beq \label{eq207}
[ \hat \partial_{i} , \hat y^{j} ] = \delta_{i}^{\,\,\,j} \, , \quad 
\{ \hat\partial_{\alpha},\teh^{\beta} \} = \delta_{\alpha}^{\,\,\,\beta} \, , 
\quad \ldots
\eeq
If nothing else is said $\partial_{i}$ is the derivative according to the 
coordinate $y^{i}$.
When $\cab$ is invertible, the fermionic derivatives $\partial_{\alpha}$ are 
internal operations in the algebra $\hat\mathcal{A}$ and they are given 
by\footnote{
$p(\hat F) = \left\{ \begin{array}{ll}
                         0: & \hat F \,\,\mathrm{bosonic} \\ 
                         1: & \hat F \,\,\mathrm{fermionic} 
                     \end{array} \right.$}
\beq \label{eq208}
\hat \partial_{\alpha} \hat F(\hat x,\teh,\tehb) = C_{\alpha\beta}^{-1} 
\left( \teh^{\beta} \hat F - (-1)^{p(\hat F)} \hat F \teh^{\beta} \right) \, ,
\eeq
which leads to the classical relation
\beq \label{eq209}
\{ \hat \partial_{\alpha} , \hat \partial_{\beta} \} = 0 \, . 
\eeq

\subsection{Symmetries}

\noindent
The algebra (\ref{eq201}) is covariant under the group of classical 
supertranslations parameterized by $(\hat a,\hat\xi,\!\bar{\,\hat\xi)}$   
\beqa \label{eq20new10}
\hat x'^{m} &=& \hat x^{m} + \hat a^{m} +i\hat\te\sigma^{m}\,\bar{\!\hat\xi} - 
i\hat\xi\sigma^{m}\,\bar{\!\hat\te} \, ,\nonumber \\
\hat\te_{\alpha}' &=& \hat\te_{\alpha} + \hat\xi_{\alpha} \, , \\
\,\bar{\!\hat\te}_{\dot\alpha}' &=& \,\bar{\!\hat\te}_{\dot\alpha} + \,
\bar{\!\hat\xi}_{\dot\alpha} \, . \nonumber
\eeqa
which is generated by the complex charges $\hat Q_{\alpha}$, 
$\,\bar{\!\hat Q}_{\dot\alpha}$ and the four momentum $\hat P_{m}$.
On the deformed chiral superspace (\ref{eq205}) the generators have the form 
\beqa
\hat P_{m} &=& i \hat\partial_{m} \, , \\
\hat Q_{\alpha} &=& \hat \partial_{\alpha} \, , \label{eq2010} \\
\!\!\bar{\,\,\hat Q_{\dot\alpha}} &=& - \!\!\bar{\,\,\hat\partial_{\dot\alpha}}
+ 2i (\teh\sigma^{m})_{\dot\alpha} \hat \partial_{m} \, ,  \label{eq2011}
\eeqa 
and satisfy following deformed supersymmetry algebra
\beqa
[ \hat P_{m},\hat P_{n} ] &=& [ \hat P_{m},\hat Q_{\alpha} ] ~=~ [ \hat P_{m},
\,\bar{\!\hat Q}_{\dot\beta} ] ~=~ 0 \, , \\
\{ \hat Q_{\alpha},\hat Q_{\beta} \} &=& 0\, ,  \\
\{ \hat Q_{\alpha},\,\bar{\!\hat Q}_{\dot\beta} \} &=& 2 \, 
\sigma^{m}_{\alpha\dot\beta} \, \hat P_{m} \, ,   \label{eq2012} \\
\{ \,\bar{\!\hat Q}_{\dot\alpha} ,\,\bar{\!\hat Q}_{\dot\beta} \} &=& 4\cab
\sigma^{m}_{\alpha\dot\alpha}\sigma^{n}_{\beta\dot\beta} \hat P_{m}
\hat P_{n}  \, . \label{eq2013}
\eeqa

\subsection{Star product}

\noindent
It is convenient to use the star product formulation of the algebra. This 
means that we use the commutative coordinates and functions but replace the 
ordinary product by the star product. The star product on this deformed 
superspace is 
\beqa 
F(\te) *  G(\te) &=&  
\exp\left(-\frac{1}{2} \cab  \partial_{\alpha}^{F} \partial_{\beta}^{G} \right)
\, F(\te)\, G(\te)  \nonumber \\
&=&  FG - \frac{1}{2} (-1)^{p(F)}  \cab \partial_{\alpha} F
 \, \partial_{\beta} G - \det \cab \frac{\partial}{\partial(\te\te)} 
F \frac{\partial}{\partial(\te\te)} G \, .\quad \label{eq2021}
\eeqa
$\partial_{\alpha}^{F}$ act only on $F$ and $\partial_{\beta}^{G}$ act only on
$G$, e.g.
\beq \label{eq2022}
\partial_{\beta}^{G} (FG) = (-1)^{p(F)} \partial_{\beta} G \, .
\eeq
Also, similar to the case of the Weyl-Moyal star product \cite{Weyl,Moyal}
one can show that 
\beq \label{eq2023}
\int d^{2} \te  F(\te)* G(\te) = \int d^{2} \te  F(\te) G(\te) \, , 
\eeq
which correspond to the fact that the difference
\beq \label{eq2024}
F(\te)* G(\te) -  F(\te) G(\te) = \partial_{\alpha} (\cdots)^{\alpha} 
\eeq
is a total Grassmann derivative not surviving the Grassmann integration.

The star product (\ref{eq2021}) is invariant under $Q$ and therefore we 
expect it to be a symmetry of the space. However, since $\bar Q$ depends 
explicitely on $\te$, it is clear that the star product is not invariant under
$\bar Q$. Therefore, $\bar Q$ is not a symmetry of the noncommutative space. 
Since half of the $N=(\frac{1}{2},\frac{1}{2})$ supersymmetry is broken, we 
can refer to the unbroken $Q$ supersymmetry as $N=(\frac{1}{2},0)$ 
supersymmetry.

\subsection{Covariant derivatives and (anti)chiral superfields}

\noindent
On the deformed chiral superspace the covariant derivatives also have the 
standard expressions 
\beqa
D_{\alpha} &=& \partial_{\alpha} + 2i (\sigma^{m}\teb)_{\alpha} \partial_{m} 
\, , \label{eq2014} \\
\bar D_{\dot\alpha} &=& - \bar\partial_{\dot\alpha} \, , \label{eq2015} 
\eeqa 
and satisfy
\beqa
\{ D_{\alpha},\bar D_{\dot\alpha} \} &=& -2i \sigma^{m}_{\alpha\dot\alpha} 
\partial_{m} \, ,\label{eq2016} \\
\{ D_{\alpha}, D_{\beta}\} &=& \{\bar D_{\dot\alpha},\bar D_{\dot\beta} \} ~=~
0 \, , \label{eq2017}\\
\{D_{\alpha},Q_{\beta}\} &=& \{\bar D_{\dot\alpha}, Q_{\beta}\} ~=~ \{ 
D_{\alpha},\bar Q_{\dot\beta} \} ~=~ \{ \bar D_{\dot\alpha},\bar Q_{\dot\beta}
\} ~=~ 0 \, . \label{eq2018}
\eeqa
Note that the change to the chiral superspace (\ref{eq205}) was needed since 
otherwise $D_{\alpha}$ would not be a derivation with respect to the star
product. Furthermore, because of the relations (\ref{eq2018}) we can use the 
covariant derivatives as in the classical case to define chiral and 
antichiral superfields. Chiral superfields are defined to satisfy 
$\,\bar{\!\hat D}_{\dot\alpha} \hat\Phi =0$ and antichiral satisfy 
$\hat D_{\dot\alpha} \bar{\hat\Phi}=0$, respectively. In components it is
\beqa
\hat\Phi (\hat y, \teh) &=& \hat A(\hat y) + \sqrt{2} \teh \hat\psi(\hat y) + 
\teh\teh \hat F(\hat y) \, , \label{eq2019} \\ 
\bar{\hat\Phi} (\bar{\hat y},\tehb) &=& \bar{\hat A}(\bar{\hat y}) + \sqrt{2} 
\teh \,\bar{\!\hat\psi}(\bar{\hat y}) + \teh\teh \bar{\hat F}(\bar{\hat y}) 
 \label{eq2020} 
\eeqa
where $\!\!\!\bar{\,\,\,\hat y^{m}}=\hat y^{m} - 2i\teh\sigma^{m}\tehb$. 
The vector superfield will be considered in the next section.

Two chiral superfields are multiplied using the star product (\ref{eq2021}). 
Clearly, the result is a function of $y$ and $\te$ and therefore it is a chiral
superfield. Two antichiral superfields of the form (\ref{eq2020}) can be 
multiplied as
\beqa 
\bar\Phi_{1} (\bar y_{1},\teb) * \bar\Phi_{2} (\bar y_{2},\teb) &=& 
 \exp \left( 2\teb\teb \cij \frac{\partial}{\partial\bar y_{1}^{i}} 
\frac{\partial}{\partial\bar y_{2}^{j}} \right) \, \bar\Phi_{1} 
(\bar y_{1},\teb) \,\bar\Phi_{2} (\bar y_{2},\teb) 
\bigg|_{y_{1},y_{2}\, \to\, y} \nonumber \\
&=& \bar\Phi_{1}(\bar y,\teb) \bar\Phi_{2} (\bar y,\teb) + 2\teb\teb\cij
\frac{\partial}{\partial\bar y^{i}} \bar\Phi_{1}(\bar y,\teb)  
\frac{\partial}{\partial\bar y^{j}} \bar\Phi_{2} (\bar y,\teb) \, .\qquad
\label{eq2025} 
\eeqa
The result is an antichiral superfield.

\subsection{Gauge theory and restriction of the gauge group}

\noindent
Consider the noncommutative gauge transformation of a chiral superfield 
$\hat\Phi$ 
\beq  \label{eq2026}
\delta_{\Lambda}\Phi = -i \Lambda * \Phi \, , 
\eeq
with the Lie algebra valued noncommutative gauge parameter 
$\hat\Lambda\!=\!\hat\Lambda_{a} T^{a}$  and $\hat{\bar D} \hat\Lambda=0$ in 
order to preserve chirality. $T^{a}$ are generators of the appropriate gauge 
group and form the Lie algebra 
\beq \label{eq2027}
[T^{a},T^{b}] = i f^{ab}_{\,\,\,c} T^{c} \, .
\eeq
The commutator of two gauge transformations has the same form as in the 
classical case
\beq \label{eq2028}
\delta_{\Lambda} \delta_{\Sigma} - \delta_{\Sigma} \delta_{\Lambda} = 
\delta_{i[\Lambda\stackrel{*}{,}\Sigma]} \, ,
\eeq
but the commutator 
\beq \label{eq2029}
[\Lambda\stackrel{*}{,}\Sigma] = \frac{1}{2} \{\Lambda_{a}
\stackrel{*}{,}\Sigma_{b}\} [T^{a},T^{b}] + \frac{1}{2} [\Lambda_{a}
\stackrel{*}{,}\Sigma_{b}] \{T^{a},T^{b}\} 
\eeq
only closes into the Lie algebra if the gauge group under consideration is 
$U(N)$. This was already shown in the same way in \cite{Terashima:0306}
and in the context of instanton calculus in \cite{Britto:0311}.
Thus in this setting gauge theories with gauge groups $SU(N)$ can not
be considered. However, using Seiberg-Witten map we can consider $SU(N)$ or 
arbitrary groups.  

The $N=(\frac{1}{2},0)$ supersymmetric $U(N)$ gauge theory is defined as 
follows \cite{Seiberg:0305,Araki:0307}. The gauge symmetry acts on the real 
Lie algebra valued vector superfield infinitesimally as 
\beq \label{eq2030}
\delta_{\Lambda} V = -i \bar{\Lambda} * e_{*}^{V} + i e_{*}^{V} * \Lambda \, ,
\eeq
where the noncommutative exponential function is defined as
\beq \label{eq2031}
e^{F}_{*} ~=~ \sum_{n=0}^{\infty} \frac{1}{n!} \,(F)^{n}_{*} ~=~ 1 + F + 
\frac{1}{2}\, F*F + \frac{1}{6}\, F*F*F + \ldots
\eeq
and $\Lambda$ and $\bar\Lambda$ are matrices of chiral and antichiral 
superfields respectively. The noncommutative chiral and antichiral field 
strenghts  
\beqa \label{eq2032}
W_{\alpha} &=& -\frac{1}{4} \bar D* \bar D \,\, e_{*}^{- V} * D_{\alpha} * 
e_{*}^{V} \, ,  \label{eq2033} \\
\bar W_{\dot\alpha} &=& \frac{1}{4}  D*D \,\, e_{*}^{V} * \bar D_{\dot\alpha} 
* e_{*}^{- V} \, ,  \label{eq2033} 
\eeqa
transform under (\ref{eq2030}) as
\beqa 
\delta_{\Lambda} W_{\alpha} &=& i[ W_{\alpha} \stackrel{*}{,}  \Lambda] \, , 
\label{eq2034} \\
\delta_{\Lambda} \bar W_{\dot\alpha} &=& i[\bar W_{\dot\alpha} 
\stackrel{*}{,}\bar \Lambda] \, , \label{eq2035}
\eeqa 

Fixing the gauge freedom by a non(anti)commutative counterpart of the 
Wess-Zumino gauge, the vector superfield is reduced to 
\beqa
V (y,\te,\teb) &=& -\,\te\sigma^{m}\teb v_{m}(y) + i \te\te\teb\bar\lambda 
(y) - i \teb\teb\te^{\alpha} \Big( \lambda_{\alpha} (y) + \frac{1}{2} 
\varepsilon_{\alpha\beta} C^{\beta\gamma} \sigma^{m}_{\gamma\dot\gamma}
 \bar\lambda^{\dot\gamma} v_{m}  \Big) \nonumber \\
&&+\, \frac{1}{2} \te\te\teb\teb\left( d(y) - i\partial_{m} v^{m} (y) \right) 
\, , \label{eq2036}  
\eeqa
where, following \cite{Seiberg:0305}, $\teb\teb\te$ term is modified so that 
the standard gauge transformation rule follows for component fields.
Then the remaining infinitesimal gauge symmetry (\ref{eq2030}), which 
preserves the gauge choice (\ref{eq2036}) is 
\beqa
\Lambda (y,\te) &=& - \alpha (y) \, , \label{eq2037}  \\
\bar\Lambda (\bar y,\bar\te) &=& - \alpha (\bar y) - i \teb\teb\cij
( \partial_{i} \alpha  v_{j} ) (\bar y) \, .\label{eq2038}
\eeqa 

We also couple matter systhem by introducing a set of (anti)chiral superfields 
transforming in appropriate representations of the gauge group, i. e. the 
(anti)fundamental one
\beq  \label{eq2039}
\delta_{\Lambda} \Phi = -i\Lambda *\Phi \, , \qquad \delta_{\Lambda} \bar \Phi
= i\bar\Phi *\bar\Lambda \, .
\eeq  
Again, to ensure the standard gauge transformations of the component fields, we
have to modify the antichiral field (\ref{eq2018}) as \cite{Araki:0307}:
\beq  \label{eq2040}
\bar\Phi (\bar y,\teb) = \bar A(\bar y) + \sqrt{2} \te \bar\psi(\bar y) + 
\te\te \left(\bar F(\bar y) + i \cij\partial_{i} \bar A v_{j} (\bar y) + 
\frac{1}{4} \cij\bar A v_{i}v_{j}(\bar y) \right) \, .
\eeq
Then the $N=(\frac{1}{2},0)$ supersymmetric gauge theory is described by the 
following Lagrange density
\beq   \label{eq2041}
\mathcal{L} = \mathcal{L}_{YM} + \mathcal{L}_{int}
\eeq
where
\beqa
\mathcal{L}_{YM} &=& Tr\,\, W^{\alpha}* W_{\alpha} \big|_{\te\te}  + 
Tr\,\, \bar W_{\dot\alpha}* \bar W^{\dot\alpha} \big|_{\teb\teb} \, ,
 \label{eq2042} \\
\mathcal{L}_{int} &=&  \bar\Phi * e_{*}^{eV}* \Phi \big|_{\te\te\teb\teb} \, .
\label{eq2043} 
\eeqa
The component expansions of the Lagrange densities (\ref{eq2042}) and 
(\ref{eq2042}) are \cite{Seiberg:0305,Araki:0307} 
\beqa
\mathcal{L}_{YM} &=& \mathcal{L}_{YM}(C=0) - 2i\cij Tr\, f_{ij} \bar\lambda
\bar\lambda + \frac{|C|^{2}}{2} Tr\,\, (\bar\lambda\bar\lambda)^{2} \, , 
\label{eq2044} \\
\mathcal{L}_{int} &=& \mathcal{L}_{int}(C=0) \nonumber \\
&&+ \frac{i}{2}\, \cij f_{ij} \bar A F - 
\frac{\sqrt{2}}{2} \cab\sigma^{m}_{\alpha\dot\alpha} \bar\lambda^{\dot\alpha}
\psi_{\beta} \mathcal{D}_{m} \bar A - \frac{|C|^{2}}{16} \bar A \bar\lambda
\bar\lambda F \, , \label{eq2045}
\eeqa
where
\beqa
f_{ij} = \partial_{i} v_{j} - \partial_{j} v_{i} + \frac{i}{2} [v_{i},v_{j}]\,
 , \label{eq2046}\\
 \mathcal{D}_{m} \bar A = \partial_{m}\bar A + \frac{i}{2} v_{m}  \bar A  \, . 
\label{eq2047}
\eeqa

\subsection{Construction of the Seiberg-Witten map in terms of component 
fields}

To determine the Seiberg-Witten map we start with the Seiberg-Witten equation 
for the gauge superfield $\hat V$ 
\beq \label{eq2048}
\hat V(V) + \hat\delta_{\hat\Lambda} \hat V(V) = \hat 
V(V + \delta_{\Lambda} V) \, . 
\eeq
The hat indicate the dependence on the classical fields. To simplify matters 
we consider the abelian case and furthermore choose the Wess-Zumino gauge in 
which the superfields $\hat V$, $\hat \Lambda$ and $\bar{\hat\Lambda}$ have the
forms (\ref{eq2036}), (\ref{eq2037}) and (\ref{eq2038}). 

Following the procedure described in \cite{Seiberg,Mikulovic:0310} we expand 
first the superfields in the non(anti)commutative parameter $\cab$ 
\beqa
\hat V &=& V + V^{'} (V,\cab) + o(C^{2}) \label{eq2049} \, , \\ 
\hat\Lambda &=& \Lambda + \Lambda^{'}(\Lambda,V,\cab) + o(C^{2}) 
\label{eq2050} \, ,
\eeqa
and solve the Seiberg-Witten equation (\ref{eq2048}) perturbatively order by 
order in the noncommutativity parameter $\cab$. To zeroth order we get the 
classical gauge transformations. To first order we get 
\beq
V^{'} - V^{'}(V + \delta_{\Lambda} V) + i(\Lambda^{'} - \bar\Lambda^{'}) = 
\frac{i}{2} \, \cab \, \partial_{\alpha} V \partial_{\beta} (\Lambda + 
\bar\Lambda) \, , \label{eq2051}
\eeq  
From this equation we can read off the Seiberg-Witten equations for the 
component fields. With the assumption that the noncommutative component fields
depend only on their classical counterparts and the classical gauge field 
$v_{m}$, we get the following Seiberg-Witten equations for component fields 
to first order in $\cab$  
\beqa
v_{m}'(v) - v_{m}'(v + \delta_{\alpha} v) -2 \, \partial_{m} \alpha'(\alpha,v)
 &=& 0 \, , \label{eq2052} \\   
\lambda'(v,\lambda) - \lambda'(v + \delta_{\alpha} v,\lambda) &=& 0 \, , 
\label{eq2053} \\
\bar\lambda'(v,\bar\lambda) - \bar\lambda'(v + \delta_{\alpha} v,\bar\lambda)
&=& 0 \, , \label{eq2054} \\
d'(v,d) - d'(v + \delta_{\alpha} v,d) &=& 0 \, , \label{eq2055}
\eeqa
since in the abelian case $\delta_{\alpha}\lambda=\delta_{\alpha}
\bar\lambda=\delta_{\alpha} d=0$.

These equations have the same form as the homogenous Seiberg-Witten equations
in the case of canonically deformed superspace 
\cite{Putz:0205,Dayi:0309,Mikulovic:0310}. However, we can not simply take the
known homogenous solutions from canonically deformed superspace since the mass
dimension of the noncommutativity parameters differs. 

With the help of dimensional analysis we can see that there are no local 
solutions of the equations (\ref{eq2052})-(\ref{eq2055}). We will demonstrate
this briefly with the Seiberg-Witten map for the gauge field $v_{m}$. The mass 
dimesions of $v_{m}$, the bosonic derivative $\partial_{m}$ and the 
noncommutativity parameter $\cij$ are 
\beq \label{eq2056}
[ v_{m} ] = [ \partial_{m} ] = 1 \, ,  \quad [ \cij ] = -1 \, .
\eeq
For the Seiberg-Witten map for $v_{m}'$ we make the ansatz
\beq  \label{eq2057}
v_{m}' (C,v) = \cij F_{ijm} (v,\partial v, \ldots) \, .
\eeq
whereas $F_{ijm} (v,\partial v, \ldots)$ represent all possible local terms 
one can build from the gauge field $v_{m}$ and its derivatives with the given
index structure. Thus the mass dimension of $F_{ijm} (v,\partial v, \ldots)$ is
\beq  \label{eq2058}
[F_{ijm} (v,\partial v, \ldots)] \geq 3 \, ,
\eeq
and we get the wrong mass dimension for the field $v_{m}'$. From equation 
(\ref{eq2049}) it is obvious that $v_{m}'$ has the same mass dimension as 
$v_m$. The same can be shown for $\alpha'$, $\lambda'$, $\bar\lambda'$ and 
$d'$.

By considering equation (\ref{eq2051}), it is more naturale to assume that the
noncommutative component fields $v_{m}'$, $\lambda'$ etc. could depend on 
every classical field of the gauge supermultiplet and not only on their 
classical counterpart e.g. 
\beq \label{eq2059}
v_{m}'=v_{m}'(v,\lambda,\bar\lambda,d,\partial v,\partial\lambda,\ldots) \, .
\eeq 
In this case there exist local Seibeerg-Witten maps with right mass dimension
only for the field $d'$ e.g.
\beq \label{eq2060}
d' = a \cij \lambda\sigma_{ij} \lambda \, ,
\eeq
where $a$ is a arbitrary constant. 

One interesting question is if the nonlocal Seiberg-Witten maps lead to an 
action which is invariant under $N=\frac{1}{2}$ supersymmetry. This question is
automatically answered by solving the Seiberg-Witten equation in terms of 
superfields. For this we will apply the method developed by Wess and 
collaborators in \cite{Madore:0012,Jurco:0006,Jurco:0102,Jurco:0104} to 
determine the Seiberg-Witten maps for the superfield case.

\subsection{Construction of the Seiberg-Witten map in terms of superfields}

We consider  again the noncommutative gauge transformations of the chiral 
and antichiral matter fields (\ref{eq2039}), but with enveloping algebra 
valued gauge parameters $\hat\Lambda$ and $\bar{\hat\Lambda}$, e.g.
\beq \label{eq2061}
\hat\Lambda = \Lambda_{a} T^{a} + \Lambda_{ab}^{'} :T^{a} T^{b}: + 
\Lambda_{abc}^{''} :T^{a} T^{b} T^{c}: + \dots \, ,
\eeq
where $\Lambda^{'}$ is linear in $\cij$, $\Lambda^{''}$ is quadratic in $\cij$,
etc.The dots indicate that we have to sum over a basis of the vector space 
spanned by the homogenous polynomials in the generators $T^{a}$ of the Lie 
algebra. 

The commutator of two transformations (\ref{eq2039}) is certainly enveloping 
algebra valued. Hence we can use arbitrary Lie groups but the price we seem 
to have to pay is an infinite number of gauge parameters and an infinite number
of gauge fields. 

To avoid this problem we define new gauge transformations, where all these  
infinitely many gauge parameters depend just on the classical gauge parameter 
$\Lambda$ or $\bar\Lambda$ respectively, the classical gauge field $V$ and on 
their derivatives. We assume moreover that all superfields considered (e.g. 
$\hat\Phi$, $\hat V$) depend on their classical counterparts, the classical 
gauge field $V$ and on their derivatives. This dependence, which we call 
Seiberg-Witten map, will be denoted by $\hat\Lambda(\Lambda,V)$, 
$\bar{\hat\Lambda}(\bar\Lambda,V)$, $\bar{\hat\Phi}(\Phi,V)$ and $\hat V(V)$. 

The gauge transformations for the noncommutative chiral and antichiral matter 
fields and the noncommutative gauge field now have the form 
\beqa 
\delta_{\Lambda} \hat\Phi(\Phi,V) &=& -i \hat\Lambda(\Lambda,V)  *  \hat\Phi
(\Phi,V)\, , \label{eq2062} \\
\delta_{\Lambda} \bar{\hat\Phi}(\bar\Phi,V) &=& i \bar{\hat\Phi}(\Phi,V) *  
\bar{\hat\Lambda}(\bar\Lambda,V) \, , \label{eq2063} \\  
\delta_{\Lambda} \hat V(V) &=& -i \bar{\hat\Lambda}(\bar\Lambda,V) * 
e_{*}^{\hat V(V)} + i e_{*}^{\hat V(V)} * \hat{\Lambda}(\Lambda,V) \, . 
\label{eq2064}
\eeqa 
Since equation (\ref{eq2028}), called consistency condition in
\cite{Jurco:0104}, involves solely the gauge parameters, it is convenient to 
base the construction of the Seiberg-Witten map on it and on the corresponding
condition for the gauge parameters of the antifundamental representation. In 
a second step the remaining Seiberg-Witten maps for the matter fields and the 
gauge field can be computed from the equations (\ref{eq2062})-(\ref{eq2064}).
 
The procedure in the abelian case is the following. As was mentioned, we start
with the consistency conditions which have the following form in the abelian 
case
\beqa 
(\delta_{\Lambda} \delta_{\Sigma} - \delta_{\Sigma} \delta_{\Lambda}) \, 
\hat\Phi(\Phi,V) = 0 \, , \label{eq2065} \\
(\delta_{\Lambda} \delta_{\Sigma} - \delta_{\Sigma} \delta_{\Lambda}) \, 
\bar{\hat\Phi}(\bar\Phi,V) = 0 \, . \label{eq2066}
\eeqa  
With equations (\ref{eq2062}) and (\ref{eq2063}) we get more explicitly
\beqa 
i \delta_{\Lambda} \hat\Sigma(\Sigma,V) - i \delta_{\Sigma} \hat\Lambda
(\Lambda,V) + \hat\Sigma(\Sigma,V) * \hat\Lambda(\Lambda,V) - \hat\Lambda
(\Lambda,V) * \hat\Sigma(\Sigma,V) = 0 \, , \label{eq2067} \\
i \delta_{\Lambda} \bar{\hat\Sigma}(\Sigma,V) - i \delta_{\Sigma} \bar{\hat
\Lambda}(\Lambda,V) + \bar{\hat\Sigma}(\Sigma,V) * \bar{\hat\Lambda}(\Lambda,V)
- \bar{\hat\Lambda}(\Lambda,V) * \bar{\hat\Sigma}(\Sigma,V) = 0 \, . 
\label{eq2068}
\eeqa
The variation $\delta_{\Lambda}\hat\Sigma(\Sigma,V)$ refers to the 
$V$-dependence of $\hat\Sigma(\Sigma,V)$ and the gauge transformation of the 
supersymmetric abelian gauge field V 
\beq \label{eq2069}
\delta_{\Lambda} V = i (\Lambda - \bar\Lambda) \, .
\eeq
We now expand the consistency conditions and the gauge transformations 
(\ref{eq2062})-(\ref{eq2064}) using the expansions (\ref{eq2049}), 
(\ref{eq2050}), the corresponding expansions for $\bar{\hat\Lambda}$, 
$\hat\Phi$, $\bar{\hat\Phi}$ and the expanded star product (\ref{eq2021}). From
these expanded equations we can then determine the Seiberg-Witten maps order by
order in $\cij$ for all considered superfields. To first order in $\cij$ we 
get the following equations 
\beqa
\delta_{\Lambda} \Sigma^{'}(\Sigma,V) - \delta_{\Sigma} \Lambda^{'}(\Lambda,V)
&=& i \cab \partial_{\alpha}\Lambda\partial_{\beta}\Sigma \, , \label{eq2070}\\
\delta_{\Lambda} \bar\Sigma^{'}(\bar\Sigma,V) - \delta_{\Sigma} \bar
\Lambda^{'}(\bar\Lambda,V) &=& i \cab \partial_{\alpha}\bar\Lambda
\partial_{\beta}\bar\Sigma \, , \label{eq2071}\\
\delta_{\Lambda} \Phi^{'}(\Phi,V) + i \Lambda\Phi^{'}( \Phi,V) + i 
\Lambda^{'}(\Lambda,V) \Phi &=& \frac{i}{2} \cab \partial_{\alpha} \Lambda
\partial_{\beta} \Phi \label{eq2072} \\
\delta_{\Lambda} \bar\Phi^{'}(\bar\Phi,V) - i \bar\Lambda\bar\Phi^{'}(\bar
\Phi,V) + i \bar\Lambda^{'}(\bar\Lambda,V) \bar\Phi &=& \frac{i}{2} \cab 
\partial_{\alpha} \bar\Lambda \partial_{\beta} \bar\Phi \label{eq2073} \\
\delta_{\Lambda} V^{'} (V) - i\Lambda^{'}(\Lambda,V) + i\bar\Lambda^{'}
(\bar\Lambda,V) &=& \frac{i}{2} \cab\partial_{\alpha} (\Lambda + \bar\Lambda) 
\partial_{\beta} V \, . \label{eq2074}
\eeqa
We will now look for solutions of these equations.

\subsection{Seiberg-Witten map for the gauge parameters}

For the gauge parameter $\Lambda{'}$ there exist a local, chiral and 
$N\!=\!(\frac{1}{2},0)$ supersymmetric Seiberg-Witten map. It is 
\beq \label{eq2075}
\Lambda^{'}(\Lambda,V) = -\,\frac{1}{2} \cab \partial_{\alpha} \Lambda 
\partial_{\beta} MV \, ,
\eeq 
where $M$ is the $N\!=\!(\frac{1}{2},0)$ chiral projector \cite{Mikulovic:0310}
\beq \label{eq2076}
M = \frac{1}{16\Box} \bar D^{2} D^{\alpha} M_{\alpha} \, ,
\eeq
with
\beq \label{eq2077}
M_{\alpha} (y) = -\partial_{\alpha} + 2i  \sigma^{m}_{\alpha\dot
\alpha} \teb^{\dot\alpha}  \partial_{m} \, ,
\eeq
For the gauge parameter $\bar\Lambda{'}$ however, there does not exist a local,
antichiral and $N\!=\!(\frac{1}{2},0)$ supersymmetric solution. This can be 
shown using dimensional analysis.  

The right hand side of the consistency condition (\ref{eq2071}) is linear in 
each of the classical superfields $\bar\Lambda$ and $\bar\Sigma$. All terms in 
the ansatz for $\bar\Lambda^{'}$ which would contain powers of $V$ can 
therefore solve only the homogeneous consistency condition because of 
(\ref{eq2069}). Hence we make an ansatz for $\bar\Lambda^{'}$ only linear in 
$V$ without loss of generality. Moreover $\bar\Lambda^{'}$ has to be linear in 
the classical gauge parameter $\bar\Lambda$ and by definition linear in $\cij$ 
or $\cab$ respectively. In order to preserve the $N\!=\!(\frac{1}{2},0)$ 
supersymmetry we may only use the bosonic derivatives $\partial_{m}$, the 
spinor derivative $\partial_{\alpha}$, the spinor coordinate $\teb$ and the 
covariant spinor derivatives (\ref{eq2014}) and (\ref{eq2015}). The mass 
dimensions of these objects are
\beqa
[ \bar\Lambda^{'} ] = [ \bar\Lambda ] = [ V ] = 0 \, , \quad [\cab]=-1 \, , 
\quad \partial_{m} = 1 \, \nonumber \\
\quad [\partial_{\alpha}]=\frac{1}{2} \, ,\quad [\teb]=-\frac{1}{2} \, , \quad
[D_{\alpha}]=[\bar D_{\dot\alpha}]=\frac{1}{2} \, .\label{eq2078}
\eeqa
It is not hard to see that there is only one term built out of this objects
with mass dimension zero and which is local, antichiral, preserve the 
$N\!=\!(\frac{1}{2},0)$ supersymmetry and have appropriate index contraction. 
It is
\beq \label{eq2079}
\bar\Lambda^{'} = a\cab \partial_{\alpha} \bar\Lambda  D^{2} \teb\teb V \, .
\eeq
where $a$ is a constant. It is obvious that there is no choice of this 
constant such that (\ref{eq2079}) is a solution of the equation (\ref{eq2071}).
Thus we have shown that for $\bar\Lambda^{'}$ there is no local and antichiral
solution which preserve the $N\!=\!(\frac{1}{2},0)$ supersymmetry. 

Using the same method one could show that even if we give up antichirality 
there is no local solution of equation (\ref{eq2071}) which preserve the 
$N\!=\!(\frac{1}{2},0)$ supersymmetry. If we give up the 
$N\!=\!(\frac{1}{2},0)$ supersymmetry there exists a local and antichiral 
expression, namely
\beq \label{eq2080}
\bar\Lambda^{'}(\Lambda,V) = \frac{1}{2} \cab \partial_{\alpha} \bar\Lambda 
\partial_{\beta} \bar  M V \, ,
\eeq  
which solves the equation  (\ref{eq2071}). $\bar M$ is the 
$N\!=\!(0,\frac{1}{2})$ antichiral projector \cite{Mikulovic:0310}
\beq \label{eq2081}
\bar M = \frac{1}{16\Box} D^{2} \bar D_{\dot\alpha} \bar M^{\dot\alpha} \, ,
\eeq
with
\beq \label{eq2082}
\bar M^{\dot\alpha} (y) = - \bar \partial^{\dot\alpha} + 4i \bar\sigma^{m\,
\dot\alpha\alpha} \te_{\alpha} \partial_{m} \, .
\eeq

Nonlocal Seiberg-Witten maps for  $\Lambda^{'}$ and $\bar\Lambda^{'}$ are 
\beqa
\Lambda^{'}(\Lambda,V) &=& -\,\frac{1}{2} \cab \partial_{\alpha} \Lambda 
\partial_{\beta} PV \, ,  \label{eq2083} \\
\bar\Lambda^{'}(\bar\Lambda,V) &=& \quad\frac{1}{2} \cab \partial_{\alpha} \bar
\Lambda \partial_{\beta} \bar P V \, .  \label{eq2084} 
\eeqa
where $P$ and $\bar P$ are the covariant chiral and antichiral superfield 
projectors
\beq
P = \frac{\bar D^{2} D^{2}}{16 \Box} \, , \qquad \bar P = \frac{ D^{2} \bar 
D^{2}}{16 \Box}  \, . \label{eq2085}  
\eeq
These solutions preserve the $N\!=\!(\frac{1}{2},0)$ supersymmetry and are 
chiral and antichiral, respectively. The component expansion of these 
solutions in the Wess-Zumino gauge are 
\beqa
\Lambda^{'} &=& 0 \, , \label{eq2086} \\
\bar\Lambda^{'} &=& \teb\teb \cij\partial_{i} \alpha \frac{\partial_{j}}{\Box}
\left( d - i\partial^{m} v_{m} \right) \, .  \label{eq2087}   
\eeqa
We will now look for solutions of the equations (\ref{eq2072})-(\ref{eq2074}).

\subsection{Seiberg-Witten maps for the matter and gauge fields}

It is not hard to find solutions of the equations (\ref{eq2072})-(\ref{eq2074})
which correspond to the nonlocal solutions  for $\Lambda^{'}$  (\ref{eq2081}) 
and $\bar\Lambda^{'}$  (\ref{eq2082}). They are
\beqa
\Phi^{'} (\Phi,V) &=& -\frac{1}{2}\cab \partial_{\alpha} \Phi \partial_{\beta}
PV \, , \label{eq2088} \\
\bar\Phi^{'} (\bar\Phi,V) &=& \,\,\,\,\, \frac{1}{2}\cab \partial_{\alpha} 
\bar\Phi \partial_{\beta} \bar PV \, , \label{eq2089} \\
V^{'} (V) &=& -\frac{1}{2}\cab \partial_{\alpha} V \partial_{\beta} PV + 
\frac{1}{2}\cab \partial_{\alpha} V  \partial_{\beta} \bar PV  -\frac{1}{2}\cab
\partial_{\alpha} PV  \partial_{\beta} \bar PV \, . \quad \label{eq2090}
\eeqa 
The component expansion of these superfields in Wess-Zumino gauge are
\beqa
\Phi^{'}(y,\te) &=& A^{'}(y) + \sqrt{2}\te\psi'(y) \, , \label{eq2091}  \\
\bar\Phi^{'}(y,\teb) &=& \teb\teb \bar F^{'}(y) \, , \label{eq2092} \\
V^{'}(y,\te,\teb) &=& \teb\bar\chi'(y) + \teb\teb M(y) - \te\sigma^{m}\teb 
v_{m}'(y) - i\teb\teb\te \lambda'(y) \, , \label{eq2093}    
\eeqa
where the Seiberg-Witten maps for the component fields are
\beqa
A^{'}(\psi,\bar\lambda) &=& \frac{1}{\sqrt{2}} \,\cij\, \psi\sigma_{j}
\frac{\partial_{i}}{\Box} \bar\lambda \, , \label{eq2094} \\
\psi_{\alpha}'(F,\bar\lambda) &=& \cij \left( \sigma_{j} 
\frac{\partial_{i}}{\Box}\bar\lambda \right)_{\alpha} F \, , \label{eq2095} \\
\bar F^{'}(A,v,d) &=& - \cij \partial_{i} \alpha \frac{\partial_{j}}{\Box}
\left( d - i \partial^{m} v_{m} \right) \, , \label{eq2096} \\
\bar\chi^{'\dot\alpha}(v,\bar\lambda,d) &=& \cij \left( 
\bar\sigma^{m}\sigma_{j} \frac{\partial_{i}}{\Box} \bar\lambda 
\right)^{\dot\alpha} \left( \frac{\partial_{m}}{\Box} \left( d-i\partial^{n}
v_{n} \right) - \frac{1}{2} v_{m} \right) \, , \label{eq2097} \\
M^{'}(v,\lambda,\bar\lambda,d) &=& -\frac{i}{2} \cij \left( \lambda \sigma_{j} 
\frac{\partial_{i}}{\Box}\bar\lambda  + \frac{1}{2} v_{j} 
\frac{\partial_{i}}{\Box} \left( d-i\partial^{n}v_{n} \right) \right) \, ,
\label{eq2098} \\
v_{m}'(\bar\lambda) &=& \frac{i}{2} \cij \frac{\partial_{i}}{\Box}\bar\lambda
\bar\sigma_{j}\sigma_{m}\bar\lambda \, , \label{eq2099} \\
\lambda_{\alpha}'(v,\bar\lambda,d) &=& \frac{i}{2} \cij \left( \sigma_{j} 
\frac{\partial_{i}}{\Box} \bar\lambda ( d-i\partial^{n}v_{n} ) + \frac{1}{2}
\sigma_{j}\bar\lambda\frac{\partial_{i}}{\Box} ( d-i\partial^{n}v_{n} ) \right)
\, . \label{eq20100}
\eeqa
As expected, this maps are nonlocal and did not coincide with the original ones
proposed by Seiberg and Witten when the superpartner fields are set to 
zero.

The Seiberg-Witten maps for the matter and gauge fields which correspond to the
solutions (\ref{eq2075}) and (\ref{eq2080}) are obtained by replacing $P$ with
$M$ and $\bar P$ with $\bar M$ in equations (\ref{eq2088})-(\ref{eq2090}). 
In Wess-Zumino gauge these maps simplify to
\beqa
\Phi^{'} (\Phi,V) &=& 0 \, , \label{eq20101} \\
\bar\Phi^{'} (\bar\Phi,V) &=& \frac{1}{2}\cab \partial_{\alpha} 
\bar\Phi \partial_{\beta} \bar MV \, , \label{eq20102} \\
V^{'} (V) &=& \frac{1}{2}\cab \partial_{\alpha} V  \partial_{\beta} \bar MV  
\, , \quad \label{eq20103}
\eeqa
since in this gauge $\partial_{\alpha} MV=0$. It is clear that these solutions 
breke tha $N\!=\!(\frac{1}{2},0)$ supersymmetry because of 
$\{ Q,\bar M \}\!\neq\! 0$ . Furthermore this solutions are nonlocal as well. 
For this reasons we will not consider them further. 

We come now to the Seiberg-Witten maps for the field strengths and the $U(1)$
Yang-Mills Lagrange density.

\subsection{Seiberg-Witten map for the field strengths and Yang-Mills action}

The expansion of the gauge field (\ref{eq2049}) induce a expansion of the 
fieldstrengths in terms of the non(anti)commutative parameter $\cab$ 
\beqa
\hat W_{\alpha} &=& W_{\alpha} + W_{\alpha}^{'} (V,\cab) + o(C^{2}) \, , 
\label{eq20104} \\
\bar{\hat W}_{\dot\alpha} &=& \bar W_{\dot\alpha} + \bar W_{\dot\alpha}^{'}
(V,\cab) + o(C^{2})  \, . \label{eq20105}
\eeqa
The Seiberg-Witten map for the field strengths can be determined from equations
(\ref{eq2032}) and (\ref{eq2033}). In the Wess-Zumino gauge we get to first 
order in $\cab$
\beqa
W_{\alpha}^{'} (V) &=& -\frac{1}{4} \bar D^{2} \left( D_{\alpha} V^{'} + 
\frac{1}{2} C^{\beta\gamma} \partial_{\beta} D_{\alpha} V
\partial_{\beta} V \right) \, , \label{eq20106} \\
\bar W_{\dot\alpha}^{'} (V) &=& -\frac{1}{4} D^{2} \left( D_{\dot\alpha} V^{'} 
- \frac{1}{2} C^{\beta,\gamma} \partial_{\beta} D_{\dot\alpha} V
\partial_{\beta} V \right)   \, .  \label{eq20107}
\eeqa
It is obvious that these Seiberg-Witten maps are also nonlocal. 
With (\ref{eq2093}) we obtain for the expansion of $W_{\alpha}^{'}$ and 
$\bar W_{\dot\alpha}^{'}$ in terms of clasical component fields 
\beqa
W_{\alpha}^{'} (y,\te) &=& -i\lambda_{\alpha}' - i(\sigma^{m}\partial_{m}\bar
\chi')_{\alpha} - i (\sigma^{m}\bar\sigma^{n}\te)_{\alpha}\partial_{m} v_{n}'
\nonumber \\ 
&&+ \frac{i}{2}\, \cij (\sigma_{j}\bar\lambda)_{\alpha} v_{i} + \frac{1}{2}\,
\cij (\sigma_{ij}\te)_{\alpha} \bar\lambda\bar\lambda \, ,\label{eq20108} \\
\bar W_{\dot\alpha}^{'} (\bar y,\teb) &=& -i(\teb\bar\sigma^{m}\sigma^{n}
)_{\dot\alpha} \partial_{m} v_{n}' + \teb\teb\bigg( \Box\bar\chi_{\dot\alpha}' 
+ (\partial_{m}\lambda'\sigma^{m})_{\dot\alpha} \nonumber \\
&&+\, \cij\partial_{j}
(\bar\lambda_{\dot\alpha}v_{i}) - \frac{1}{2} \cij\partial_{m}(v_{i}
\bar\lambda\bar\sigma^{m}\sigma_{j})_{\dot\alpha} \bigg) \, , \label{eq20109} 
\eeqa
where $\bar\chi'$, $v_{m}'$ and $\lambda'$ are given in (\ref{eq2097}), 
(\ref{eq2099}) and  (\ref{eq20100}).

The Yang-Mills Lagrange density expanded up to first order in $\cij$ is
\beqa
\mathcal{L}_{YM } &=& \hat W^{\alpha} * \hat W_{\alpha} \big|_{\te\te}  +
\bar{\hat W}_{\dot\alpha} * \bar{\hat W}^{\dot\alpha} \big|_{\teb\teb} 
\nonumber \\
&=& \mathcal{L}_{YM}(C=0) + 2 W^{\alpha} W_{\alpha}^{'} \big|_{\te\te}
+ 2 \bar W_{\dot\alpha}^{'} \bar W^{\dot\alpha} \big|_{\teb\teb} + o(C^{2})
\, . \label{eq20110}
\eeqa 
With the equations (\ref{eq20108}) and (\ref{eq20109}) we obtain up to total
derivatives
\beqa
W^{\alpha} W_{\alpha}^{'} \big|_{\te\te} &=& i\bar\lambda\Box\bar\chi' + i
\partial_{m}\lambda'\sigma^{m}\bar\lambda + id\partial^{m}v_{m}' - 
\frac{1}{2}\, f_{ab}^{SD} f^{'ab} \nonumber \\
&&- \frac{i}{2}\, \cij \partial_{m}\bar\lambda\bar\sigma^{m}\sigma_{j}\bar
\lambda v_{i} - \frac{i}{2}\, \cij f_{ij} \bar\lambda\bar\lambda \, ,
\label{eq20111} \\
\bar W_{\dot\alpha}^{'} \bar W^{\dot\alpha} \big|_{\teb\teb} &=& i\bar\lambda
\Box\bar\chi' + \partial_{m}\lambda'\sigma^{m}\bar\lambda  + id\partial^{m}
v_{m}' - \frac{1}{2}\, f_{ab}^{ASD} f^{'ab} \nonumber \\
&&+ \frac{i}{2}\, \cij \bar\lambda\bar\sigma^{m}\sigma_{j}\partial_{m}\bar
\lambda v_{i} - \frac{i}{2}\, \cij f_{ij} \bar\lambda\bar\lambda \, ,
\label{eq20112}
\eeqa
where $f_{ab}'=\partial_{a} v_{b}' - \partial_{b} v_{a}'$ and $f_{ab}^{SD}$ and
$f_{ab}^{ASD}$ are the selfdual and antiselfdual field strengths, respectively.
It is obvious that $\mathcal{L}_{YM}$ (\ref{eq20110}) is nonlocal due to the 
fields $\bar\chi'$, $\lambda'$ and $v_{m}'$. It is also worth noticing, that 
this Lagrange density has also the same term as the Lagrange density 
(\ref{eq2043}).

The Lagrange density (\ref{eq20110}) differs from the Lagrange density   
(\ref{eq2043}) by one local and three nonlocal terms. Thus, the Lagrange 
density (\ref{eq20110}) is nonlocal, as expected.

\section{$N\!=\!(\frac{1}{2},\frac{1}{2})$ deformed Euclidean superspace}

We consider now the following deformed superspace 
\cite{Ferrara:0307,Saemann:0401}:
\beqa \label{eq20117}
&& [\hat x^{i} ,\hat x^{j} ] = \teb\teb\cij \, , \nonumber \\
\hat\mathcal{R}:
&&  [\hat x^{i} , \teh^{\alpha} ] = -i \cab \sigma^{m}_{\beta\dot\beta} 
\!\bar{\,\hat\theta^{\dot\beta}} \, , \quad [\hat x^{i} , 
\!\bar{\,\hat\theta^{\dot\beta}} ] = 0 \, , \\
&& \{ \teh^{\alpha} , \teh^{\beta} \} = \cab \, , \quad
\{ \!\bar{\,\hat\theta^{\dot\alpha}} , \!\bar{\,\hat\theta^{\dot\beta}} \} = 
\{ \teh^{\alpha} , \!\bar{\,\hat\theta^{\dot\alpha}} \} = 0 \nonumber \, ,
\eeqa
where the equations (\ref{eq202})-(\ref{eq204}) still hold. The noncommutative
functions, fields and derivatives are again defined as in 
(\ref{eq206})-(\ref{eq209}). The algebra (\ref{eq20117}) is also covariant 
under the group of classical supertranslations (\ref{eq20new10}). 

It is convenient to use the chiral coordinates $y^{m}=x^{m} + i\te\sigma^{m}
\teb$ instead of $x^{m}$. One may readily check that the (anti)commutation 
relations (\ref{eq20117}) become
\beqa \label{eq20118}
&& [\hat y^{i} ,\hat y^{j} ] = 4\teb\teb\cij \, , \nonumber \\
\hat\mathcal{R}:
&&  [\hat y^{i} , \teh^{\alpha} ] = -2i \cab \sigma^{m}_{\beta\dot\beta} 
\!\bar{\,\hat\theta^{\dot\beta}} \, , \quad [\hat y^{i} , 
\!\bar{\,\hat\theta^{\dot\beta}} ] = 0 \, , \\
&& \{ \teh^{\alpha} , \teh^{\beta} \} = \cab \, , \quad
\{ \!\bar{\,\hat\theta^{\dot\alpha}} , \!\bar{\,\hat\theta^{\dot\beta}} \} = 
\{ \teh^{\alpha} , \!\bar{\,\hat\theta^{\dot\alpha}} \} = 0 \nonumber \, .
\eeqa
On this deformed chiral superspace the generators of supertranslation and the 
covariant deivatives have the same form and the same anticomutation relations
as in sections 2.1 and 2.3. The chiral and anichiral superfields are also 
defined in standard way, namely $\,\bar{\!\hat D}_{\dot\alpha} \hat\Phi =0$ and
$\hat D_{\dot\alpha} \bar{\hat\Phi}=0$.

The star product is given by 
\beq \label{eq20119}
F(\te) *  G(\te) =  \exp\left(-\frac{1}{2} \cab D_{\alpha}^{F} D_{\beta}^{G} 
\right) \, F(\te)\, G(\te) \, 
\eeq
where $D_{\alpha}^{F}$ acts only on $F$ and $D_{\beta}^{G}$ acts only on $G$, 
e.g.
\beq \label{eq20120}
D_{\beta}^{G} (FG) = (-1)^{p(F)} D_{\beta} G \, .
\eeq
Although the star product (\ref{eq20119}) preserves the $N\!=\!(\frac{1}{2},
\frac{1}{2})$ supersymmetry, it does not allow to define chiral superfields 
which form subalgebras of the star product. For this reson the deformation 
(\ref{eq20118}) leads for the Wess-Zumino model to the same Lagrangian as the 
deformation (\ref{eq205}), preserving only the $N\!=\!(\frac{1}{2},0)$ 
supersymmetry \cite{Ferrara:0307,Saemann:0401}. 

The gauge theory can be constructed following section 2.4. The equations 
(\ref{eq2026})-(\ref{eq2035}) still hold. Moreover, in \cite{Saemann:0401} it 
was demonstrated that also the Yang-Mills actions on both superspaces are 
equivalent. Because of equation (\ref{eq2029}) we can consider again only gauge
theories with $U(N)$ gauge groups. To consider gauge theories with $SU(N)$ 
gauge groups we have to determine the Seiberg-Witten map.

\subsection{Construction of the Seiberg-Witten maps in terms of superfields}

The steps in constructing the Seiberg-Witten maps are the same as in section 
2.6. The equations (\ref{eq2061})-(\ref{eq2069}) did not change. We consider 
here again the abelian case. With the star product (\ref{eq20119}) we get the
following equations from which the Seiberg-Witten maps to first order in 
$\cij$ can be determined:
\beqa
\delta_{\Lambda} \Sigma^{'}(\Sigma,V) - \delta_{\Sigma} \Lambda^{'}(\Lambda,V)
&=& i \cab D_{\alpha}\Lambda D_{\beta}\Sigma \, , \label{eq20121} \\
\delta_{\Lambda} \bar\Sigma^{'}(\bar\Sigma,V) - \delta_{\Sigma} \bar
\Lambda^{'}(\bar\Lambda,V) &=& 0 \, , \label{eq20122}\\
\delta_{\Lambda} \Phi^{'}(\Phi,V) + i \Lambda\Phi^{'}( \Phi,V) + i 
\Lambda^{'}(\Lambda,V) \Phi &=& \frac{i}{2} \cab D_{\alpha} \Lambda
D_{\beta} \Phi \, , \label{eq20123} \\
\delta_{\Lambda} \bar\Phi^{'}(\bar\Phi,V) - i \bar\Lambda\bar\Phi^{'}(\bar
\Phi,V) + i \bar\Lambda^{'}(\bar\Lambda,V) \bar\Phi &=& 0\, ,\label{eq20124} \\
\delta_{\Lambda} V^{'} (V) - i\Lambda^{'}(\Lambda,V) + i\bar\Lambda^{'}
(\bar\Lambda,V) &=& \frac{i}{2} \cab D_{\alpha} \Lambda  D_{\beta} V \, . 
\label{eq20125}
\eeqa
We will now look for solutions of these equations.

\subsection{Seiberg-Witten maps for gauge parameters}

First of all we have to find solutions for $\Lambda^{'}$ and $\bar \Lambda{'}$
since the Seiberg-Witten maps for the matter fields and the gauge field depend 
on them. For $\bar \Lambda{'}$ we can simply take
\beq \label{eq20126}
\bar \Lambda{'} (\Lambda,V) = 0 \, ,
\eeq
which is a solution of the equation (\ref{eq20122}). The other solutions of 
the equation (\ref{eq20122}) either breake the 
$N\!=\!(\frac{1}{2},\frac{1}{2})$ supersymmetry or locality. This is obvious 
from dimensional analysis. The mass dimension of $\bar \Lambda^{'}$ is zero
(\ref{eq2078}). The Seiberg-Witten map for $\bar \Lambda^{'}$ contains at 
least $\bar\Lambda$, $\cab$ or $\cij$ respectively and the gauge invariant 
field strengths $W_{\alpha}$ or $\bar W_{\dot\alpha}$. The mass dimension of 
these three objects is
\beq \label{eq20127}
\left[ C\Lambda W \right] = \frac{1}{2} \, .
\eeq 
In order to get the right mass dimension we have to use operators with negativ
mass dimension. One posibility is to use $\teb\teb$, e.g.
\beq \label{eq20128}
\bar \Lambda^{'} (\bar\Lambda,W) \sim  \cab \teb\teb\bar\Lambda D_{\alpha}
W_{\beta} \, ,
\eeq
but these solutions possess only $N\!=\!(\frac{1}{2},0)$ symmetry. The other 
posibility is to use nonlocal operators, e.g
\beq \label{eq20129}
\bar \Lambda^{'} (\bar\Lambda,\bar W) \sim \cab \sigma^{m}_{\alpha\dot\alpha}
\bar D^{\dot\alpha} \bar\Lambda \frac{\partial_{m}}{\Box} W_{\beta} \, . 
\eeq 

It is not hard to see that the Seiberg-Witten equations for $\bar \Lambda{''}$,
 $\bar \Lambda{'''}$, ..., are of homogenous type like equation 
(\ref{eq20122}), e.g.
\beq \label{eq20130}
\delta_{\Lambda} \bar\Sigma^{''}(\bar\Sigma,V) - \delta_{\Sigma} \bar
\Lambda^{''}(\bar\Lambda,V) = 0 \, .
\eeq  
This is the case because of the star product (\ref{eq20119}). Thus we can take 
the trivial solution for all these fields and get
\beq \label{eq20131}
\bar{\hat\Lambda} = \bar\Lambda \, ,
\eeq
which means that the antichiral gauge parameter remains undeformed. 

We come now to the solutions for $\Lambda^{'}$. From equation (\ref{eq20121})
it is obvious that $\Lambda^{'}$ must contain the nonchiral term 
$D_{\alpha}\Lambda$. Thus there does not exist a chiral Seiberg-Witten map
for $\Lambda^{'}$ which solves equation (\ref{eq20121}). A nochiral, local and 
$N\!=\!(\frac{1}{2},\frac{1}{2})$ supersymmetric solution is
\beq  \label{eq20132}
\Lambda^{'} \left( \Lambda,V \right) = -\frac{1}{2} \cab D_{\alpha} \Lambda 
D_{\beta} V \, .
\eeq
The component expansion of this solution is
\beq \label{eq20133}
\bar\Lambda^{'} = \cij \teb\teb \left( iv_{i} \partial_{j} \alpha + 
\te\sigma_{j}\bar\lambda\partial_{i}\alpha  \right) \, .
\eeq

We can find easily further solutions for $\Lambda^{'}$ using the chiral 
projectors $P$ (\ref{eq2085}) and $M$ (\ref{eq2076}). These are
\beqa
\Lambda^{'} \left( \Lambda,V \right) &=& -\frac{1}{2} \cab D_{\alpha} \Lambda 
D_{\beta} PV \, , \label{eq20134} \\
\Lambda^{'} \left( \Lambda,V \right) &=& -\frac{1}{2} \cab D_{\alpha} \Lambda 
D_{\beta} MV \, . \label{eq20135}
\eeqa
These solutions are not only nonchiral but also nonlocal, since
\beqa
PV(y) &=& \frac{1}{2\Box} \left( d-i\partial^{m}v_{m} \right) -\te\sigma^{m}
\frac{\partial_{m}}{\Box} \bar\lambda \, , \label{eq20136} \\
MV(y) &=& -\frac{1}{\Box} \left( d-i\partial^{m}v_{m} \right) \, . 
\label{eq20137} 
\eeqa 
Moreover, the solution (\ref{eq20137}) possesses just the 
$N\!=\!(\frac{1}{2},0)$ supersymmetry.

Now we want to find solutions for the equations (\ref{eq20123})-(\ref{eq20125})
which correspond to the solution (\ref{eq20126}) and to the solutions 
(\ref{eq20132}), (\ref{eq20134}) and (\ref{eq20135}), respectively. 
It is clear that in this case equations (\ref{eq20123}) and (\ref{eq20125}) 
become
\beqa
\delta_{\Lambda} \Phi^{'}(\Phi,V) + i \Lambda\Phi^{'}( \Phi,V)  &=& 
\frac{i}{2} \cab D_{\alpha} \Lambda D_{\beta} \Phi \, , \label{eq20138} \\
\delta_{\Lambda} V^{'} (V) - i\Lambda^{'}(\Lambda,V)  &=& \frac{i}{2} \cab 
D_{\alpha} \Lambda  D_{\beta} V \, . \label{eq20139}
\eeqa

\subsection{Seiberg-Witten maps for matter and gauge fields}

The Seiberg-Witten map for $\bar\Phi^{'}$ which corresponds to the solution 
(\ref{eq20126}) is simply
\beq \label{eq20140}
\bar\Phi^{'} (\bar\Phi,V) = 0 \, .
\eeq
For the field $\bar\Phi^{'}$ it can be shown in exactly the same manner as for
the field $\bar\Lambda^{'}$ in previous section, that we can take the trivial
Seiberg-Witten map to all orders in $\cab$ 
\beq \label{eq20141}
\bar\Phi^{''} = \bar\Phi^{'''} = \bar\Phi^{''''} = \dots = 0 \, , 
\eeq
and get
\beq \label{eq20142}
\bar{\hat\Phi} = \bar \Phi \, \, ,
\eeq  
which means that the antichiral matter field remain undeformed.

Among the solutions for $\Phi^{'}$ and $V^{'}$ which correspond to the 
solution (\ref{eq20132}) for $\Lambda^{'}$, there are also the trivial 
solutions
\beq \label{eq20143}
\Phi^{'} (\Phi,V) = V^{'} (V) = 0 \, .
\eeq 
Thus, in this case to first order in $\cab$ only the gauge parameter field is
deformed whereas all other fields remain undeformed. 

The Seiberg-Witten maps for $\Phi^{'}$ and $V^{'}$ which correspond to the 
solution (\ref{eq20134}) are
\beqa  
\Phi^{'} \left( \Phi,V \right) &=& -\frac{1}{2} \cab D_{\alpha} \Phi 
D_{\beta} PV \, , \label{eq20144} \\
V^{'} \left( V \right) &=& -\frac{1}{2} \cab D_{\alpha} V 
D_{\beta} PV \, , \label{eq20145}
\eeqa
and which correspond to the solution (\ref{eq20135}) are
\beqa  
\Phi^{'} \left( \Phi,V \right) &=& -\frac{1}{2} \cab D_{\alpha} \Phi 
D_{\beta} MV \, , \label{eq20146} \\
V^{'} \left( V \right) &=& -\frac{1}{2} \cab D_{\alpha} V 
D_{\beta} MV \, . \label{eq20147}
\eeqa
respectively. It is obvious that all this solutions are nonlocal due to 
equations (\ref{eq20136}) and (\ref{eq20137}). Furthermore, the Seiberg-Witten
maps (\ref{eq20144}) and  (\ref{eq20146}) for the matter field $\Phi^{'}$ are
nonchiral and the later possesses only the 
$N\!=\!(\frac{1}{2},0)$ supersymmetry.

\section{Conclusions}

We have considered Seiberg-Witten maps for superfields on the  
$N\!=\!(\frac{1}{2},0)$ and the $N\!=\!(\frac{1}{2},\frac{1}{2})$ deformed 
superspaces. We have shown that on the $N\!=\!(\frac{1}{2},0)$ deformed 
superspace there is no Seiberg-Witten map for antichiral superfields which is 
at the same time antichiral, local and which preserve the 
$N\!=\!(\frac{1}{2},0)$ supersymmetry. Solutions wich break this requirements
were presented.

On the $N\!=\!(\frac{1}{2},\frac{1}{2})$ deformed superspace we have shown 
that for the chiral gauge parameter, and therefore also for the chiral matter
field, there is no chiral Seiberg-Witten map. Some other possible 
Seiberg-Witten maps for the superfields were presented.
It is worth noticing that the antichiral fields remain undeformed on this 
superspace.

\section*{Acknowledgements}

I want to thank Fabian Bachmaier, Branislav Jurco and Ivo Sachs for useful 
discussions. I also wish to thank Julius Wess for drawing my attention to
noncommutative superspaces.


\end{document}